\documentclass[journal]{IEEEtran}
\ifCLASSINFOpdf
\else
\fi
\usepackage{amsmath}
\usepackage{amssymb,xspace,intmacros,subfigure,amsfonts}
\usepackage{epsfig,epsf,epstopdf,graphicx}
\usepackage{color}
\newcommand\ie{i.e.\xspace}

\begin{document}
\title{Non-Coherent Direction of Arrival Estimation via Frequency Estimation}

\author{Hadi~Zayyani and
Mehdi~Korki,~\IEEEmembership{Student Member,~IEEE}~
        %and~Farrokh~Marvasti,~\IEEEmembership{Senior Member,~IEEE}% <-this % stops a space
\thanks{H. Zayyani is with the Department
of Electrical and Computer Engineering, Qom University of Technology, Qom, Iran (e-mail: zayyani@qut.ac.ir).}% <-this %
\thanks{M. Korki is with School of Software and Electrical
Engineering, Swinburne University of Technology, Hawthorn, Vic. 3122,
Australia (e-mail: mkorki@swin.edu.au).}
%\thanks{M. Korki is with School of Software and Electrical Engineering, Swinburne University of Technology, Hawthorn,
%3122 Australia (e-mail: mkorki@swin.edu.au).}

%stops a space
%\thanks{Manuscript received April 19, 2005; revised December 27, 2012.}
}


\markboth{IEEE Signal Processing Letters,~Vol.XX, No.X}%
{Shell \MakeLowercase{\textit{et al.}}:}



\maketitle

\begin{abstract}
This letter investigates the non-coherent Direction of Arrival (DOA) estimation problem dealing with the DOA estimation from magnitude only measurements of the array output. The magnitude squared of the array output is expanded as a superposition of some harmonics. Hence, a frequency estimation approach is used to find some nonlinear relations between DOAs, which results in inherent ambiguities. To solve the nonlinear equations and resolve the ambiguities, we assume a high amplitude reference target at low angles to estimate the true DOA's with no ambiguities. However, the proposed algorithm requires a large number of antenna array elements to accurately estimate the DOA's. To overcome this drawback, and to enhance the estimation accuracy, we suggest two variants of the algorithm. One is to add virtual elements in the array and the second is to integrate multiple snapshots. Simulation results show that the proposed frequency estimation-based algorithm outperforms the non-coherent GESPAR algorithm in the low signal to noise ratio (SNR) regime and it is two orders of magnitude faster than the non-coherent GESPAR algorithm.

\end{abstract}


 \ifCLASSOPTIONpeerreview
 \begin{center} \bfseries EDICS: SAM-DOAE, SAS-STAT \end{center}
 \fi
\IEEEpeerreviewmaketitle

\section{Introduction}

\IEEEPARstart{D}{irection} of arrival (DOA) estimation is a classical problem in signal processing with a variety of practical applications \cite{Greco09}\nocite{Thompson93}--\cite{Godara97}. In the literature, some conventional algorithms \cite{Johnson92}, \cite{Capon69}, \cite{Schm86} and sparsity-based algorithms \cite{Wang09}\nocite{Zheng13}\nocite{Malioutov05}\nocite{Gurbuz12}--\cite{Rossi14} have been proposed for DOA estimation. However, all these algorithms rely on the available ideal phase synchronization (at the elements of the array), which is difficult to achieve in practice. This letter focuses on the problem of non-coherent DOA estimation which is recently proposed in the literature by Kim {\it et al.} \cite{Kim15}. In this problem, the DOA estimation is performed based on magnitude only measurements of the array output. This is very effective when the phase errors are present at array element level or the phase synchronization does not exist. Phase synchronization is usually difficult to achieve, specially in phased array radars. To eliminate the sensitivity of the classical DOA estimation algorithms such as MUSIC \cite{Schm86} and ESPRIT \cite{Rich89} to phase errors, an algorithm based on phase retrieval is suggested in \cite{Kim15}. To estimate the DOAs, the authors used a modified version of a phase retrieval algorithm called GESPAR \cite{Shec14}. To resolve the inherent ambiguities of the problem, they used a reference target at low angles which is capable of solving the ambiguities for the single target scenario. In the context of multiple targets scenario, they suggested to use multiple reference targets to reduce the ambiguities. Recently,  Jiang {\it et al.} presented the exact formula for approximate maximum likelihood estimation of the problem in the case of single target scenario \cite{Jian16}. They also calculated the Cramer-Rao Bound (CRB) for the non-coherent DOA estimation in the single target scenario.

In this letter, we expand the expression of magnitude squared of the array signal output. Assuming the array element index as a time index, the signal in terms of element index can be considered as a superposition of some harmonics. Hence, a simple harmonic analysis tool such as FFT analysis can recover the above mentioned multiple harmonics. Extracting the frequency of these harmonics results in $\frac{K(K-1)}{2}$ nonlinear equations of the unknown DOAs where $K$ is the number of targets. To solve these nonlinear equations and also to resolve the ambiguity arises in these equations, we propose a simple practical solution. We use only one high amplitude reference target at low angles. In fact, we reduce the number of nonlinear equations to $K$, and thus we just use $K$ largest peaks of the spectrum. By this practical trick, we can estimate the true DOAs and resolve the ambiguity arises in the nonlinear equations. The main drawback of this approach is that it requires many array elements to form the spatial signal. To combat this drawback, we virtually add some elements to the array by simple classical sampling techniques. Moreover, one advantage of the proposed technique is that it only requires one snap shot. To further extend the proposed method to exploit the multiple snap shots and increase the accuracy of DOA estimations, we also suggest to coherently integrate the multiple snap shots similar to those methods in radar signal processing.

To resolve the ambiguities in multiple targets scenario, we propose to utilize a high amplitude reference target at low angles. This is an advantage compared to the method in \cite{Kim15}, where the multiple reference targets have been suggested intuitively to resolve the ambiguities. Simulation results show the superiority of the proposed algorithm over the non-coherent GESPAR \cite{Kim15} for the low SNR regime ($SNR<10$ dB). Simulation results also verify that the proposed algorithm is about two orders of magnitude faster than non-coherent GESPAR algorithm.

%The rest of the letter is organized as follows. Section~\ref{sec: Problem} introduces the problem formulation. The GLRT detector and the optimal quantizer design is discussed in Section~\ref{sec: GLRT}. A double GLRT detector is suggested in Section~\ref{sec: doubleGLRT}. Section~\ref{sec: clair} introduces some clairvoyant and oracle detectors. Simulation results are presented in Section~\ref{sec: Sim}. Finally, conclusions are drawn in Section~\ref{sec: con}.

\section{The problem formulation}
\label{sec: Problem}
Consider a linear array of $N$ elements at locations $z_n=nd$ where $n=0,1,...,N-1$ and $d$ is the spacing distance between two adjacent elements. Consider $K$ sources in direction angles of $\theta_k,k=1,\cdots,K$ in far-field, impinging independent narrowband signals $x_k(t),k=1,\cdots,K$ on an array in an isometric environment. For the $i$'th target, the received signal at $n$'th sensor is $x_{n,i}(t)=|x_{n,i}(t)|\mathrm{e}^{j(z_n\Psi_i(t)+\gamma_{n,i}(t))}$ where $|x_{n,i}(t)|$ is the received amplitude, $\Psi_i(t)=\frac{2\pi}{\lambda}\mathrm{\cos}(\theta_i(t))$, and $\gamma_{n,i}(t)$ is the phase error. Hence, the received signal at the $n$'th sensor is
\begin{equation}
\label{eq: model}
y_n(t)=\sum_{i=1}^K|x_{n,i}(t)|\mathrm{e}^{j(z_n\Psi_i(t)+\gamma_{n,i}(t))},
\end{equation}
where $|x_{n,i}(t)|$ are unknown amplitudes and $\gamma_{n,i}(t)$ are unknown phase errors. We aim to estimate the direction of arrivals $\theta_i$ from the magnitude only measurements (i.e. $|y_n(t)|^2$) of the array outputs. Intentionally, we do not use the classical matrix form of (\ref{eq: model}) which is found in the literature and mostly used for sparse recovery-based approaches for coherent DOA estimation \cite{Kim15}, \cite{Malioutov05}, \cite{Jaga13}.

\section{Frequency estimation-based algorithm}
\label{sec: freq}

\subsection{Basic Idea}
If we consider the magnitude squared of the array output as a function of time index, it is composed of some harmonics, each of which is related to cross terms of two available targets. Considering a specific time index and omitting some unnecessary subscripts, we have the magnitude squared signal as
\begin{equation}
a_n=|y_n|^2=|(\sum_{i=1}^K|x_{n,i}|\mathrm{e}^{j(z_n\Psi_i+\gamma_{n,i})})|^2,
\end{equation}
where the time index $t$ is omitted for simplicity. Writing $|y_n|^2=y_ny^{*}_n$, with some simple calculations, we have:
\begin{equation}
\label{eq: an}
a_n=\sum_{i=1}^K\sum_{i'=1}^K|x_{n,i}||x_{n,i'}|\mathrm{e}^{j\frac{2\pi nd}{\lambda}(\cos(\theta_i)-\cos(\theta_{i'}))}\mathrm{e}^{j(\gamma_{n,i}-\gamma_{n,i'})},
\end{equation}
where $a_n$ consists of a dc term and $\frac{K(K-1)}{2}$ harmonic terms, each of which is related to a cross term of two available targets. Each harmonic term has a normalized frequency given by
\begin{equation}
\label{eq: freqpeak}
\tilde{f}_{i,i'}=\frac{d}{\lambda}|\cos(\theta_i)-\cos(\theta_{i'})|.
\end{equation}
If we use a frequency estimation tool to find these frequencies (e.g, taking FFT and finding the peaks of magnitudes), we have $\frac{K(K-1)}{2}$ nonlinear equations such as (\ref{eq: freqpeak}) and we aim to estimate the DOAs, \ie $\theta_i$, from the nonlinear equations (\ref{eq: freqpeak}).

\subsection{Ambiguities}
If we have two targets, we have only one nonlinear equation. Hence, we are unable to accurately estimate the two DOAs. Considering one target as a reference target at a low angle similar to the technique suggested in \cite{Kim15}, we can find the DOA of the other target assuming that the DOA is in the range $[0^\circ,90^\circ]$. If the DOA of the target is in the range $[90^\circ,180^\circ]$, we can use another array with the reference target at $\theta_i=180^\circ$. Hence, for removing the ambiguity of the targets whose DOA's are in $[90^\circ,180^\circ]$, which arises due to the absolute value operator in (\ref{eq: freqpeak}), we use another array with the reference target at $\theta=180^\circ$. Therefore, after all, we restrict ourselves to DOA's in the range $[0^\circ,90^\circ]$.

If there are more than two targets the ambiguities are still not resolved. To cope with this problem, the authors in \cite{Kim15} proposed to use multiple reference targets. In this letter, we instead suggest to use only one high amplitude reference target at a low angle. If the echo from this reference target is sufficiently large, the amplitudes of the cross term between this target and other targets are larger than the others. Hence, by finding the $K$ largest peaks of the FFT spectrum we can find the corresponding DOA of the targets. We have $K$ largest peaks at
\begin{equation}
\tilde{f}_{i,ref}=\frac{d}{\lambda}|\cos(\theta_i)-\cos(\theta_{ref})|, \quad 1\le i\le K.
\end{equation}
If the reference target is assumed at a low angle (i.e., $\theta_{ref}\approx 0$), the DOAs are estimated via the following formula
\begin{equation}
\label{equ_6}
\hat{\theta}_i=\cos^{-1}(-\frac{\lambda}{d}\tilde{f}_{i,ref}+\cos(\theta_{ref}))\approx \cos^{-1}(-\frac{\lambda}{d}\tilde{f}_{i,ref}+1).
\end{equation}
In deriving (\ref{equ_6}), it is assumed that the amplitude of cross term of reference target (target 1) is larger than the amplitude of other cross terms from other non-reference targets \ie $|x_{ref}||x_i|\ge|x_k||x_m|,\forall k\neq m\ge 2$. This requires that $|x_{ref}|>\frac{|x_k|^2_{max}}{|x_m|_{min}}=\mathrm{DR}\times|x_k|_{max}$ where $\mathrm{DR}$ is the dynamic range of the return amplitudes. Therefore, if we put synthetically an strong enough reference target at low angles, we could be certain that the $K$ largest peaks in the spectrum is related to the pair of reference target and other targets.

\subsection{Virtual Array}
As we explained earlier, the main idea of the proposed frequency estimation-based algorithm is to consider the array element index as time index, or we consider the magnitude squared of the array output as a superposition of some spatial frequencies. Since the accuracy of the frequency estimation algorithms improves with the signal length (\ie $N$ number of elements), we need many elements to yield good results. Increasing the number of array elements, increases the complexity of the algorithm. However, we aim to accurately estimate DOAs with a limited number of array elements. Towards that end, we suggest to virtually add some elements to the array, which is feasible thanks to the classical sampling theorem. Consider a time-continuous magnitude squared signal as
\begin{equation}
\label{eq: at}
a(s)=\sum_{i=1}^K\sum_{i'=1}^K|x_{s,i}||x_{s,i'}|\mathrm{e}^{j\frac{2\pi sd}{\lambda}(\cos(\theta_i)-\cos(\theta_{i'}))}\mathrm{e}^{j(\gamma_{s,i}-\gamma_{s,i'})},
\end{equation}
where $a_n=a(nT)$ in (\ref{eq: an}) is a sampled version of $a(s)$ in (\ref{eq: at}) with sampling interval $T=1$. According to classical sampling theorem, if the sampling frequency is higher than two times the bandwidth, the signal can be reconstructed from the discrete samples. Now, we have a signal $a(s)$ with bandwidth $\mathrm{BW}=\frac{d}{\lambda}(\cos(\theta_i)-\cos(\theta_{i'}))_{max}$. For $d=\frac{\lambda}{2}$, it is not difficult to ascertain that the sampling theorem requisite is satisfied. So, the continuous signal $a(s)$ is exactly reconstructed from sinc interpolation which is the sum of sinc functions as follows
\begin{equation}
a(s)\approx\sum_{n=0}^{N-1}a_n\mathrm{sinc}(s-n),
\end{equation}
where the approximation is due to limited number of time samples of $a(s)$. Therefore, if the complexity allows us, we can increase the number of array elements.

\subsection{Coherent Integration for Multiple Snapshots}
Although the proposed algorithm is based on only one snapshot, to further improve the accuracy of DOA estimation or to exploit the data of multiple snapshots, we can use multiple snapshots. To decrease the effect of noise, we suggest to coherently integrate the multiple snapshots similar to radar signal processing \cite{Skol02}.
 
%So, we have the coherently integrated signal as
%\begin{equation}
%\tilde{a}(s)=\sum_{t_p}a(s,t_p)
%\end{equation}

\section{Simulation Results}
\label{sec: Sim}
This section presents the simulation results to illustrate the performance of the proposed non-coherent DOA estimation method. In all experiments, three targets ($K=3$) are present with the reference target at $\theta_{\mathrm{ref}}=0$. The two other targets are assumed to be at $\theta_i\in[10^\circ,90^\circ]$. For reference, the results of the non-coherent algorithm is compared to the coherent DOA estimation, which is solved by applying the orthogonal matching pursuit (OMP) algorithm \cite{Trop07}, \cite{Kim15}. The number of angle bins is $N_{\theta}=200$ which is uniformly spaced over $[0^\circ,180^\circ]$. The spacing between elements in the linear array is assumed to be $\frac{\lambda}{2}$. Unless otherwise stated, the phase error is $\gamma_{n,i}(t)=0$. The reference target and the unknown targets have the amplitudes $|x_{n,i}(t)|=100$ for $i=1$ and $|x_{n,i}(t)|=1$ for $i\neq 1$, respectively. Signals received from the targets are added to white complex Gaussian noise with zero mean and variance $\sigma^2_v$. Similar to \cite{Kim15}, the SNR per sensor is then defined as $\mathrm{SNR}=-10\log_{10}\sigma^2_v$. The performance metric to measure the performance is the Mean Square Error (MSE) of estimated angles of non-reference targets ($i\neq 1$) which is averaged over the number of experiments:
\begin{equation}
\mathrm{MSE}=\frac{1}{M}\sum_{r=1}^M\sum_{i=2}^K(\theta_{i,r}-\hat{\theta}_{i,r})^2,
\end{equation}
where $M$, $\theta_{i,r}$, and $\hat{\theta}_{i,r}$ are the number of independent runs, different random angles of targets, and the estimated angles of targets, respectively.

\begin{figure}[tb]
\begin{center}
\includegraphics[width=7cm]{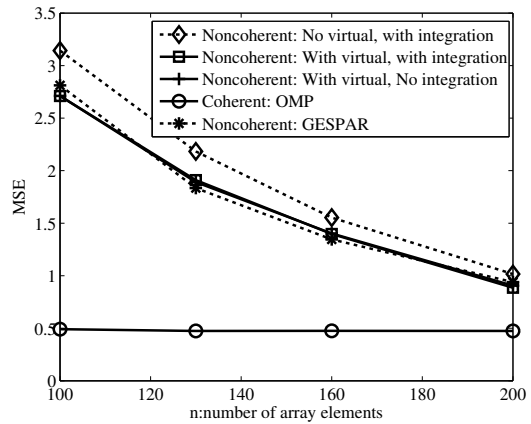}
\end{center}
\caption{MSE versus the number of array elements for non-coherent and coherent DOA estimation from the array sensor output.}
%\end{center}
\label{fig1}
\end{figure}

In the first experiment, we compare the proposed non-coherent DOA estimation method with the coherent case and non-coherent DOA estimation method proposed in \cite{Kim15}, when the number of array elements varies. The variance of noise is assumed to be $\sigma^2_v=0.04$, i.e. $\mathrm{SNR}\approx 14dB$. Figure.~\ref{fig1} shows the MSE versus the number of array elements for three versions of the proposed non-coherent DOA estimation method, the non-coherent DOA estimation method (denoted as non-coherent GESPAR) of \cite{Kim15}, and the coherent DOA estimation method. Three versions of the proposed non-coherent DOA estimation methods are without virtual array and with coherent integration (with five snapshots integrated), with virtual array (with increasing the number of array elements with a factor of two), and with both virtual array (by an increasing factor of two) and with coherent integration (with five snapshots integrated). The number of snapshots for coherent and non-coherent GESPAR is also five. It is seen that using virtual array slightly improves the performance and using coherent integration has no effect on the performance of the proposed noncoherent DOA estimation. The proposed method with virtual array performs almost as well as non-coherent GESPAR method, while its computational cost is much lower compared to non-coherent GESPAR method (See Fig. 4). It is also observed that increasing the number of array elements improves the performance of the non-coherent methods and has no effect on the performance of coherent DOA estimation method. Needless to say that the performance of the coherent DOA estimation depends on the number of angle beams $N_{\theta}$.

\begin{figure}[tb]
\begin{center}
\includegraphics[width=7cm]{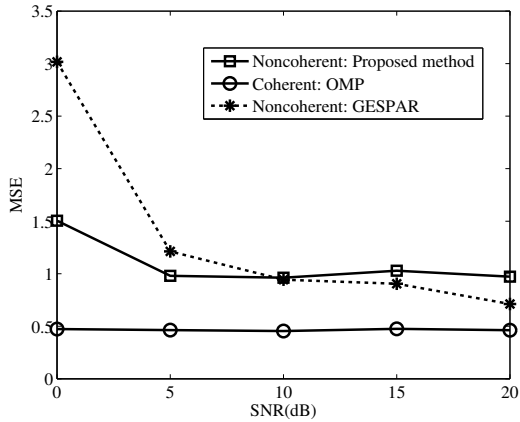}
\end{center}
\caption{MSE versus SNR for DOA estimation from the array sensor output.}
%\end{center}
\label{fig2}
\end{figure}

In the second experiment, the effect of noise on the performance of the DOA estimation is studied. The number of array elements is $n=200$. The results of MSE versus SNR for the proposed non-coherent DOA estimation (using both integration and virtual array with exactly the same settings as the first experiment), the non-coherent GESPAR method \cite{Kim15}, and the coherent DOA estimation (with $N_{\theta}=200$) are shown in Fig.~\ref{fig2}. It shows that the performance of the proposed non-coherent DOA estimation method improves by increasing the SNR for $\mathrm{SNR}<5$ dB. Also the proposed non-coherent DOA estimation outperforms the non-coherent GESPAR method for $\mathrm{SNR}<10$ dB. However, for $\mathrm{SNR}\geq10$ dB increasing the SNR has no effect on the performance and it seems that the performance is restricted by the number of array elements and the inherent limitations of the non-coherent DOA estimation. The non-coherent GESPAR method \cite{Kim15} slightly outperforms our proposed method for $\mathrm{SNR}\geq10$ dB.

%The figure also shows the negligible effect of SNR on the performance of coherent DOA estimation and the performance is probably dominated by the effect of number of angle beams.

The third experiment investigates the effect of phase errors on the performance of the DOA estimation. In this experiment, the phase errors $\gamma_{n,i}(t)$ are drawn from a uniform distribution in $[0,\mathrm{Phase Error}]$. The number of array elements is $n=200$. Figure. \ref{fig3} shows the MSE versus phase error for non-coherent and coherent DOA estimation methods. It is seen that when the phase error is approximately larger than 13 degrees, the proposed non-coherent DOA estimation method outperforms the coherent DOA estimation method. Moreover, the non-coherent GESPAR method slightly outperforms the proposed non-coherent DOA estimation method. It is also observed that phase errors have no effect on the performance of non-coherent DOA estimation methods, while the performance of coherent DOA estimation starts to collapse for phase errors larger than 25 degrees. This experiment illustrates the benefit of the non-coherent DOA estimation in comparison to the coherent case.

\begin{figure}[tb]
\begin{center}
\includegraphics[width=7cm]{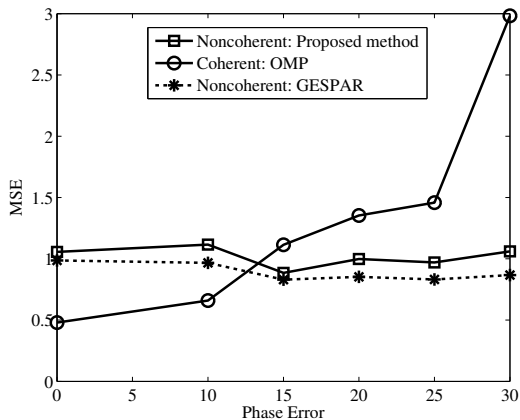}
\end{center}
\caption{MSE versus Phase error for DOA estimation from the array sensor output.}
%\end{center}
\label{fig3}
\end{figure}

The last experiment which investigates the relative complexity (or speed) of the proposed non-coherent DOA estimation, is exactly the same as the first experiment. Figure. \ref{fig3} presents the average runtime for the three versions of the proposed non-coherent DOA estimation methods, coherent method and non-coherent GESPAR method, when the number of array elements varies. The simulation has been performed in MATLAB environment using an Intel 3.10-GHz processor with 8 GB of RAM and under Windows operating system.
Figure. \ref{fig3} demonstrates the superior speed of the proposed non-coherent DOA estimation method. It is two orders of magnitude faster than the non-coherent GESPAR method, while it performs almost as well as the non-coherent GESPAR method.

\begin{figure}[tb]
\begin{center}
\includegraphics[width=7cm]{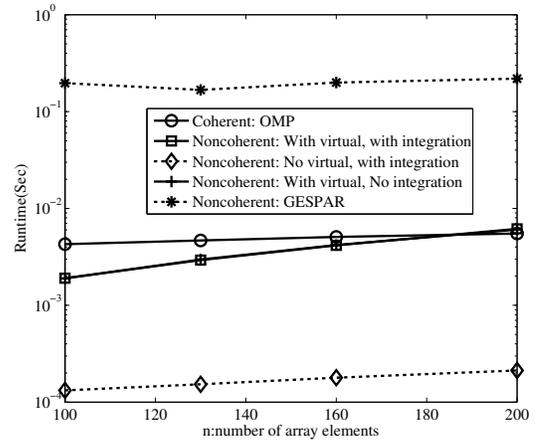}
\end{center}
\caption{Comparison of the mean runtime of the proposed non-coherent method, non-coherent GESPAR method, and coherent method for different number of  array elements.}
%\end{center}
\label{fig3}
\end{figure}

\section{Conclusion}
\label{sec: con}
In this letter, a new algorithm for the problem of non-coherent DOA estimation is proposed. It is essentially based on the harmonic estimation of the magnitude squared of the array sensor output. Considering the magnitude of the array elements as a function of element indexes, this function is composed of some harmonics where the frequency of each harmonic is related to the DOAs by nonlinear equations. By assuming the existence of a strong (or high amplitude) reference target at a low angle (e.g, zero angle), the nonlinear equations can be solved with no ambiguity and hence DOA's can be estimated. Two other ideas for improving the accuracy of the frequency estimation-based algorithm are presented. The first is to use a virtual array to increase virtually the number of array elements and the second is to use coherent integration for utilizing the data of multiple snapshots simultaneously. Simulation results show the better performance of the proposed non-coherent method than the non-coherent GESPAR for low SNR regime while non-coherent GESPAR slightly outperforms the proposed algorithm for high SNR regime. Moreover, simulation results show that the proposed frequency estimation-based algorithm is about two orders of magnitude faster than the non-coherent GESPAR method.

\section*{Acknowledgment}
Hadi Zayyani would like to thank Hojatollah Zamani for his helpful discussions and suggestions. Mehdi Korki would like to thank Haley Kim, for providing the MATLAB code of non-coherent GESPAR algorithm \cite{Kim15}.

\ifCLASSOPTIONcaptionsoff
  \newpage
\fi

\end{document}